\documentstyle[twocolumn,aps,amsfonts,amssymb,psfig,prl]{revtex}

\begin{document}
\draft
\preprint{{\bf ETH-TH/98-??}}

\title{Marginal Pinning of Quenched Random Polymers}

\author{D.A. Gorokhov and  G. Blatter} 

\address{Theoretische Physik, ETH-H\"onggerberg, CH-8093
Z\"urich, Switzerland}

\twocolumn[ 
\date{\today}
\maketitle
\widetext
\vspace*{-1.0truecm}
\begin{abstract}
\begin{center}
\parbox{14cm} {An elastic string embedded in 3D space and 
subject to a short-range correlated random potential exhibits 
marginal pinning at high temperatures, with the pinning length
$L_c(T)$ becoming exponentially sensitive to temperature. Using a
functional renormalization group (FRG) approach we find $L_c(T)
\propto \exp[(32/\pi)(T/T_{\rm dp})^3]$, with $T_{\rm dp}$ the
depinning temperature. A slow decay of disorder correlations as it
appears in the problem of flux line pinning in superconductors
modifies this result, $\ln L_c(T)\propto T^{3/2}$.}

\end{center}

\end{abstract}
] 
\vspace{-0.4truecm}

\narrowtext 

Elastic manifolds subject to a disorder potential \cite{Halpin} define
a rich and challenging problem in modern classical statistical
mechanics, with numerous applications in condensed matter physics
\cite{Halpin,Blatter,IV,Gruner}. Fundamental questions arise regarding
the relevance of the disorder \cite{Larkin} and its quantitative
effects on the manifold's quenched fluctuations on short \cite{Larkin}
and large scales \cite{Halpin}.  Such static results then determine,
via scaling arguments, the creep type motion of the elastic manifolds
under the action of an external force, rendering these studies also
relevant in the context of applied physics, e.g., plastic deformations
in metals \cite{IV} or dissipation in superconductors
\cite{Blatter}. Classifying the problems through the dimensionality
$d$ of the manifold and the number $n$ of transverse dimensions, the
($1+n$)-dimensional problem describing strings moving in $n$
directions, also known as the random polymer problem, is particularly
interesting \cite{HH,HHF}. Here, depending on the number of transverse
dimensions $n$, the temperature induced fluctuations compete in
various ways with the fluctuations due to quenched disorder. In this
Letter, we analyze the competition between elasticity, disorder, and
temperature for the ($1+n$)-string problem. In particular, we present
a functional renormalization group analysis of the ($1+2$)-dimensional
problem (a string in 3D space), where the disorder turns marginal at
high temperatures; we determine the depinning temperature $T_{\rm dp}$
above which thermal smoothing leads to a collapse of the effective
disorder strength and the temperature dependent pinning length
$L_c(T)$ beyond which the disorder dominates over the elasticity.

The free energy $F({\bf u},L)= -T\ln Z({\bf u},L)$ of a string
starting and ending at the points $({\bf 0}, 0)$ and $({\bf u}, L)$
[the first (second) coordinate denotes the displacement of (position
along) the string)] is given by the partition function
\[
Z = \!\!\! \int\limits_{({\bf 0},0)}^{({\bf u},L)} \!\!\!
{\cal D}[{\bf u}^\prime] \, 
\exp\bigg\{\!\!-\frac{1}{T} \! \int_0^L \!\!\! dz^\prime 
\bigg[ \frac{\epsilon}{2} \bigg(\frac{\partial{\bf u}^\prime} 
{\partial{\bf z^\prime}}\bigg)^2 
+ U({\bf u}^\prime,z^\prime) \bigg] \bigg\},
\]
with $\epsilon$ the elasticity of the string. The disorder potential
$U({\bf u},z)$ is assumed to be a random gaussian variable with zero
mean and a correlator
\begin{equation}
\langle U({\bf u},z) U({\bf u}^\prime, z^\prime)\rangle =
K_0(|{\bf u}-{\bf u}^\prime|)\delta (z-z^\prime),
\end{equation}
with $K_0(u)$ a smooth function decaying on the scale $\xi$.  The
partition function $Z$ describes the competition between the elastic
energy $(\epsilon/2) (\partial_z{\bf u})^2 > 0$ and the disorder
potential $U({\bf u},z)$, from which the string can gain energy by
choosing minima with $U({\bf u}, z) <0$.
\begin{figure} [bt]
\centerline{\psfig{file=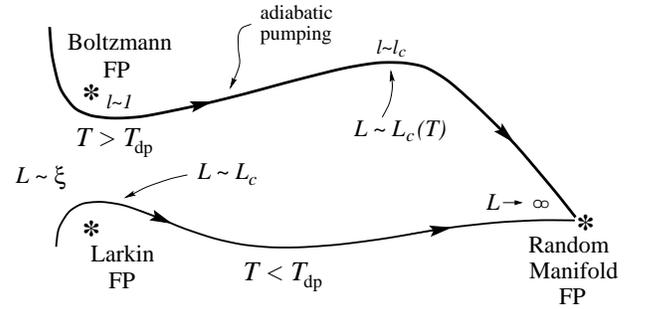,width=8cm,height=4.0cm}}
\narrowtext\vspace{4mm}
\caption{ Schematic picture of the FRG flow for the marginal case with
$n=2$: At temperatures below depinning ($T < T_{\rm dp} \approx
(\epsilon K_0(0) \xi^2)^{1/3}$) the flow proceeds through the Larkin
regime (with a fixed point characterized by the wandering exponent
$\zeta_{\rm\scriptscriptstyle L} = 3/2$) towards the random mani\-fold
fixed point with its non-trivial exponent $\zeta_2 \approx 5/8$.  At
high temperatures $T > T_{\rm dp}$ the flow starts out in the thermal
region with $\zeta_{\rm th} = 1/2$. The linear flow towards the
`Boltzmann' fixed point (BFP) rapidly erases the details in the
initial condition and produces a gaussian correlator $K_l(u) \sim g_l
\exp(-u^2 \epsilon/4\pi\xi T)$ with weight $g_l$. The non-linear terms
in the FRG produce an overall growth of the correlator while the
linear terms keep its shape to the `Boltzmann' form (adiabatic
pumping). The sharp increase of the prefactor $g_l$ at $l_c$ defines
the pinning length $L_c(T)$, beyond which the flow enters the random
manifold regime with a rapid change of the thermal exponent to the
random manifold exponent ($\zeta_{\rm th} =1/2 \rightarrow \zeta(l)
\rightarrow \zeta_2 \approx 5/8$) and a renormalization of the
temperature towards zero. The flow ends in the zero temperature random
manifold fixed point.}
\label{F1}
\end{figure}
The quantity characterizing the behavior of the string is the
displacement correlator $\langle u^2 (L)\rangle \equiv \langle[{\bf
u}(L)-{\bf u}(0)]^2\rangle$ $\propto L^{2\zeta}$ describing its
wandering with distance $L$. We distinguish between the perturbative
(Larkin \cite{Larkin}) regime at short scales $L < L_c$ and the random
manifold regime \cite{Fisher} at $L > L_c$, as well as temperatures
below and above the depinning temperature $T_{\rm dp}\approx (\epsilon
K_0(0) \xi^2)^{1/3}$, see Fig.\ \ref{F1}: For $T < T_{\rm dp}$ the
wandering exponent $\zeta_{\rm\scriptscriptstyle L} =3/2$ at small
distances $L<L_c$ (see \cite{Blatter}), while at large distances
$L>L_c$, $\zeta_n$ depends on the dimensionality of the space:
$\zeta_1 = 2/3$ is an exact result\cite{HHF} while for $n=2$ numerical
simulations \cite{AV} give a value $\zeta_2 \approx 5/8$. The
crossover length $L_c$ separating these two regimes has the form
\cite{Blatter} $L_c\approx [\epsilon^2\xi^4/K_0(0)]^{1/3}$.  At high
temperatures $T > T_{\rm dp}$ the short scale behavior is subject to
thermal smoothing, resulting in a thermal wandering with $\zeta_{\rm
th} = 1/2$, while at large scales $L > L_c(T)$ the disorder induced
line wandering characterized by $\zeta_n$ prevails. The crossover
length $L_c(T)$ increases algebraically with temperature $L_c(T)\sim
L_c \, (T/T_{\rm dp})^5$ for the ($1+1$)-case \cite{IV}, while for the
$(1+2)$-dimensional problem $L_c(T)$ is exponentially sensitive to
temperature with \cite{FV} $L_c(T) \sim A(T)\exp[C(T/T_{\rm dp})^3]$,
and $C$ a constant of order unity. It turns out that the usual
perturbative methods neither provide the value of $C$ nor that of the
prefactor $A(T)$. The exponential sensitivity of the pinning length
$L_c(T)$ to the temperature $T$ is related to the appearance of a
phase transition in dimensions $n>2$, separating a low temperature
disorder dominated phase from a high temperature thermal phase
characterized by $\zeta_{\rm th}$ \cite{Imbrie}. The $(1+2)$-problem
then corresponds to the lower critical dimension for this phase
transition and thus exhibits {\it marginal} behavior.

The main goal of the present work is to develop a consistent scheme
for calculating the pinning length $L_c(T)$ in the $(1+2)$-problem
specifying the coefficient $C$ in the exponent. While this goal can be
achieved within a one-loop calculation including three-replica terms,
it turns out that the determination of the prefactor $A(T)$ requires a
two-loop analysis (accounting for four-replica terms) which is beyond
the scope of this letter.  We first discuss perturbative methods for
calculating $L_c(T)$ at high temperatures and different dimensions
$n$. These techniques fail for the ($1+2$)-problem and we will then
turn to the more powerful renormalization group method.

{\it Perturbation Theory:} With the weak disorder providing the small
parameter $\Delta = \int d^n u\, K_0(u)$ we expand the $k$-fold
replicated Green function (see \cite{GB}) $G({\bf u}_1,\dots, {\bf
u}_k;L)$ $\equiv \langle Z({\bf u}_1,L)\dots Z({\bf u}_k,L)\rangle$
and calculate the mean squared displacement $\langle u^2(L)\rangle =
\lim_{k\rightarrow 0}$ $\int d^n u d^n u_2 \dots d^n u_k \, u^2 G({\bf
u},{\bf u}_2 \dots, {\bf u}_k;L)$. The disorder is relevant when
$\langle u^2 (L\rightarrow\infty)\rangle$ grows faster than the
thermal wandering $\langle u^2(L)\rangle_{\rm th} = (nT/\epsilon) L$
which is dominant at small lengths; the comparison of the two
expressions then defines the characteristic crossover length $L_c(T)$,
$(nT/\epsilon)\, L_c (T) \sim \langle u^2(L_c(T)) \rangle$. To lowest
order the mean squared displacement takes the form
\begin{equation}
\langle u^2(L)\rangle = \frac{nTL}{\epsilon} 
\left [1+\frac{(\Delta/T^2)(\epsilon/T)^{n/2}}{(4\pi)^{n/2}(4-n)} 
L^{1-n/2} \right];
\label{loword}
\end{equation}
for $n<2$ the disorder corrections grow more rapidly than the thermal
wandering and we obtain \cite{IV} $L_c(T) \sim L_c \,
[T/(\epsilon\Delta \xi)^{1/3}]^5$ for $n=1$. For $2<n<4$ the
contribution from the random potential goes to zero at large scales
and the disorder is irrelevant \cite{Imbrie}.  The case $n=2$ is
marginal: to lowest order the disorder correction is $L$-independent
and the next order term $\propto \Delta^2$ provides a logarithmic
correction,
\begin{equation}
\langle u^2(L)\rangle= \frac{2TL}{\epsilon}
\left[ 1+\frac{1}{8\pi} \frac{\epsilon\Delta}{T^3} +
\frac{1}{16\pi^2}\frac{\epsilon^{2}\Delta^2}{T^6}\ln (L/\xi)\right],
\label{logcorr}
\end{equation}
producing an exponential temperature dependence $L_c(T) \propto$
$\exp[CT^3/\epsilon\Delta]$ with $C$ of order unity (we compare terms
$\propto \Delta^2$ and $\propto \Delta^3$).  The arbitrariness of the
criterion prevents us from finding the coefficient in the exponent and
the prefactor and we have to develop a more systematic way in order to
deal with the problem.

{\it Renormalization group:} The basic idea is to construct the
renormalized effective correlator $K_l$ of the disorder potential
describing the behavior of the manifold on short and large
scales. With the disorder becoming irrelevant for (internal)
dimensionalities $d>4$ \cite{Larkin}, the effect of disorder is
analyzed within an $\epsilon=4-d$-expansion.  Starting from short
scales, the RG flow goes through a special point where $K_l$ becomes
singular. This point is identified with the pinning length $L_c$ of
the string \cite{Bucheli}. Its physical meaning is clear: the fact
that the correlator becomes singular at $L_c$ implies that the
perturbation theory breaks down and the manifold splits up into
independently pinned domains of size $L_c$.  By way of introduction we
briefly derive the zero-temperature pinning length $L_c$, starting
from the one-loop FRG equation \cite{Fisher,Balents}
\begin{eqnarray}
\partial_l K_l
&=&(4-d-4\zeta -n\zeta)K_l+\zeta\nabla\cdot({\bf u}K_l)\nonumber\\
&+& I [K_l^{\mu\nu}({\bf u}) K_l^{\mu\nu}({\bf u})/2
-K_l^{\mu\nu}({\bf u})K_l^{\mu\nu}({\bf 0})],
\label{frg**} 
\end{eqnarray}
with $\zeta(l)=(1/2){\partial\ln\langle u^2\rangle}/{\partial\ln L}$
the wandering exponent, $I=A_d/(2\pi)^d \epsilon^2 \Lambda^{4-d}$
($A_d={2\pi^{d/2}}/{\Gamma (d/2)}$, and $\Lambda^{-1}\sim\xi$ is the
short scale cutoff), and the indices $\mu$ and $\nu$ denote
derivatives with respect to the Cartesian coordinates $u_{\mu}$ and
$u_{\nu}$. The RG variable $l$ is related to the length $L$ via $l=\ln
{\Lambda L}$. We proceed with differentiating (\ref{frg**}) four times
with respect to $u_\kappa$.  Substituting $u=0$, the wandering
exponent $\zeta$ drops out and we obtain
\begin{equation}
\partial_l \Gamma_l = (4-d) \Gamma_l + 
[I(n+8)/3] \Gamma_l^2,
\label{der4}
\end{equation}
where $\Gamma_l={\partial^4 K_l}/{\partial u_{\kappa}^4}|_{{\bf
u}=0}$.  Integrating (\ref{der4}) we encounter a divergence at $l_c
\approx \ln[3(4-d)/I(n+8)\Gamma_0]/(4-d)$ (we assume weak disorder
with $I \Gamma_0 \ll 1$), allowing us to define the collective pinning
or Larkin length $L_c = \Lambda^{-1} \exp(l_c) \approx
[\epsilon^2\xi^4/K(0)]^{1/(4-d)}$ where simple perturbation theory
breaks down. The FRG equation (\ref{frg**}) allows us to further
characterize the Larkin regime: On short scales we can neglect the
non-linear term and an expansion of the correlator $K_l(u) \sim
\alpha(l) u^2/2$ produces the wandering exponent
$\zeta_{\rm\scriptscriptstyle L}=(4-d)/2$ at the Larkin fixed point
\cite{GiamarchiLeDoussal}. Note that, although Eq.~(\ref{frg**}) is
written in the form of an $\varepsilon$-expansion, the result for
$L_c$ is valid for any $\epsilon$ as long as we investigate the short
scale behavior, where the functional renormalization group is just a
way to sum the perturbation expansion. Indeed, one can show that in
the non-marginal situation higher order loop corrections to
Eq.~(\ref{frg**}) do not change the result for $L_c$, while in the
marginal case (e.g., for $d=4$ at $T=0$) we have to account for
two-loop corrections.

Next, we proceed to study finite temperatures, where besides a simple
diffusive term $\propto T \Delta K_l$, new three- and higher replica
terms are generated describing non-gaussian fluctuations in the
disorder potential $U$. Generalizing the disorder statistics to
include non-gaussian terms and replicating $n$ times we arrive at a
new Hamiltonian ${\cal H}_n = {\cal H}_{\rm free}+{\cal H}_{\rm int}$
with ${\cal H}_{\rm free} =\int d^d
z\sum_{\alpha}(\epsilon/2)(\partial_{\bf z}{\bf u}_{\alpha})^2$ and
\[
\frac{{\cal H}_{\rm int}}{T}
=-\int d^d z \biggl[\sum_{\alpha,\beta} 
\frac{K({\bf u}_{\alpha\beta})}{2T^2}
+\sum\limits_{\alpha,\beta,\gamma}
\frac{S({\bf u}_{\alpha\beta},{\bf u}_{\alpha\gamma})}{T^3} 
+\dots\biggr].
\]
Here, ${\bf u}_{\alpha\beta} \equiv {\bf u}_\alpha-{\bf u}_\beta$ and
$S({\bf u},\bar{\bf u})$ is the three-replica term describing
non-gaussian fluctuations in the disorder potential (the $\dots$ stand
for higher-replica terms depending on three or more variables). At
finite temperatures, a consistent analysis to one loop order requires
to include both two- and three-replica terms $K({\bf u})$ and $S({\bf
u},\bar{\bf u})$ in the FRG flow. After a second order cumulant
expansion in the two- and three-body interaction ${\cal H}_{\rm int}$,
we proceed with the standard momentum shell RG, expand the ${\bf
u}$-fields into Fourier series and perform the integration over fast
modes. The resulting equations take the form
\begin{eqnarray}
\partial_l K_l ({\bf u})
&=& (4\!-\!d\!-\!4\zeta\!-\!n\zeta)K_l+\zeta\nabla\cdot({\bf u}K_l)
+   I^{\prime} T K_l^{\mu\mu} \nonumber\\
&+& I [(1/2) K_l^{\mu\nu}({\bf u}) K_l^{\mu\nu}({\bf u})/2
    -K_l^{\mu\nu}({\bf u})K_l^{\mu\nu}({\bf 0})]\nonumber\\
&-& \!2I^{\prime}T[S_l^{\mu\mu}({\bf 0},{\bf u})
    \!+\!S_l^{{\bar{\mu}}\bar{\mu}}({\bf u},{\bf 0})
    \!-\![\partial_{\bf u}\!\cdot\!
         \partial_{\bf \bar{u}}S_l]({\bf u},{\bf u})
\nonumber\\
&+&  [\partial_{\bf u}\!\cdot\!
      \partial_{\bf \bar{u}}S_l]({\bf 0},{\bf u})
    +[\partial_{\bf u}\!\cdot\!
      \partial_{\bf \bar{u}}S_l]({\bf u},{\bf 0})],
\label{frg}\\
\partial_l S_l({\bf u},{\bf \bar{u}})
&=& (6\!-\!2d\!-\!6\zeta\!-\!2n\zeta)S_l
    +\zeta[\nabla\!\cdot\!({\bf u} S_l)
    \!+\!{\bar{\nabla}}\!\cdot\!({\bar{\bf u}}S_l)]\nonumber\\
&+& I^{\prime}T[S_l^{\mu\mu}+S_l^{\bar{\mu}\bar{\mu}}
    +\partial_{\bf u}\!\cdot\!
     \partial_{\bf \bar{u}} S_l]\nonumber\\
&+& ITK_l^{\mu\nu}({\bf u})K_l^{\mu\nu}({\bf \bar{u}})/4,
\label{ffrrgg}\\
\partial_l T_l &=&(2-d-2\zeta)T_l, 
\label{frgT}
\end{eqnarray}
where $I' = I\epsilon\Lambda^2$ and the superscripts $\mu$ ($\bar\mu$)
denote derivatives with respect to the first (second) variable in
$S_l({\bf u}, {\bf \bar{u}})$ (implicit summation over double indices
is assumed); similarly, the gradients $\nabla$ ($\bar{\nabla}$) denote
derivatives with respect to ${\bf u}$ (${\bf{\bar{u}}})$. The system
(\ref{frg}) -- (\ref{frgT}) has to be solved with initial conditions
$K_{l=0}({\bf u})=K_{0}({\bf u})$, $S_{l=0}({\bf u},{\bf \bar{u}})=0$
(the bare potential is gaussian), and $T_{l=0}=T$. Note that the
three-replica term is driven by a source $\propto T K_l^2$ and hence
is irrelevant in the $T=0$ analysis.  Furthermore, four- and higher
replica terms can be omitted within the present one-loop analysis:
with $K_l({\bf u})={\cal O}(\Delta)$ the three-replica term $S_l({\bf
u},{\bf{\bar u}})={\cal O}(\Delta^2)$, see (\ref{ffrrgg}), while the
four-replica term is driven by a term $\propto K_l^3$ and hence
$D({\bf u}_1,{\bf u}_2, {\bf u}_3)= {\cal O}(\Delta^3)$. However,
going to second-loop order the four-replica term $D_{l}$ has to be
included.

The above system (\ref{frg}) -- (\ref{frgT}) can be solved
asymptotically exactly for weak disorder and we proceed with the
analysis for the elastic string, $d=1$. On short scales the behavior
of the string is dominated by thermal fluctuations and the wandering
exponent is equal to $\zeta_{\rm th} = 1/2$; as a result, the
temperature is not renormalized, see Eq.~(\ref{frgT}). On the other
hand, on large scales $\zeta$ approaches the value $\zeta_2 \approx
5/8$ and the behavior of the string is governed by the zero
temperature fixed point. It turns out that the temperature starts
renormalizing only close to $L_c(T)$, allowing us to fix $\zeta =1/2$
in Eqs.~(\ref{frg}) -- (\ref{frgT}). The non-linear and three-replica
terms are of second order in the disorder strength and thus
unimportant at small scales $l\ll l_c$, allowing us to neglect the
terms $\propto K_l^2$ and $\propto S_l$ in (\ref{frg}) in the initial
stage of the flow. The Ansatz
\begin{equation}
K_l(u)=P_l(u)\exp[(1-n/2)l],
\label{varsubst}
\end{equation}
then maps (\ref{frg}) to the Fokker-Planck equation describing the
`time' evolution of the probability distribution function $P_l$ for an
overdamped particle in the harmonic potential $V(u)=u^2/4$ at a
`temperature' $T^\prime= T/\pi \epsilon\Lambda$, $\partial_l
P_l=\nabla\cdot({\bf u}P_l)/2 +T^\prime\Delta P_l$. At high
temperatures $T$, the distribution function $P_l(u)$ rapidly
approaches equilibrium and takes the form of a Boltzmann distribution,
\begin{equation}
P_{\rm\scriptscriptstyle B}(u)
=\Delta(4\pi T^{\prime})^{-n/2} 
\exp(-u^2/4T^\prime).
\label{incondnn}
\end{equation}
The function $P_{\rm\scriptscriptstyle B}(u)$ depends on the disorder
potential only via the integral weight $\Delta =\int d^n u \, K_0(u)$,
i.e., the evolution erases the details of the initial
condition. Depending on the dimensionality $n$, the following
situations arise (c.f., (\ref{varsubst})): For $n>2$ the solution of
(\ref{frg}) decreases with increasing $l$ and the disorder
renormalizes to zero, in agreement with the existence of a high
temperature phase when $n>2$, see \cite{Imbrie}. For $n<2$ the
correlator $K_l(u)=\exp[(1-n/2)l] P_{\rm\scriptscriptstyle B} (u)$
grows (until reaching the size of the non-linear term) and the
disorder is relevant. Comparing terms linear and quadratic in $K_l$ we
find the collective pinning length $L_c(T) \simeq
L_c[T/(\epsilon\Delta\xi)^{1/3}]^5$.  Most interesting is the marginal
case $n=2$ where we find a `Boltzmann' fixed point. In order to obtain
$L_c(T)$ we need to take into account the terms ${\cal O}(\Delta^2)$
effectively pumping additional `particles' into the system and leading
to an overall growth in the disorder strength $\Delta_l=\int d^n u\,
K_l(u)$. However, for small $\Delta_l$ the strong diffusion will relax
this `pumping' towards the `Boltzmann' distribution, i.e., the
solution of Eq.~(\ref{frg}) keeps the form
\begin{equation}
K_l({\bf u}) = g_l\, ({T^\prime}^2/I)\,\exp(-{\bf u}^2/4T^{\prime}),
\label{simplef}
\end{equation}
with the dimensionless coupling $g_l=(\pi/4) \Delta_l\epsilon/T^3$.
After diagonalizing the term $[S_l^{\mu\mu}+S_l^{\bar{\mu}\bar{\mu}}
+\partial_{\bf u}\!\cdot\!\partial_{\bf \bar{u}} S_l]$ in
(\ref{ffrrgg}) we find its Green function $G_S({\bf u},\bar{\bf
u};{\bf u}^\prime,{\bar{\bf u}}^\prime;l) =
1/12[\pi{I^\prime}T(1\!-\!e^{-l})]^2 \exp\{-[({\bf u}\!-\!{\bf
u}^\prime e^{-l/2})^2\!+ \!(\bar{\bf u}\!-\!{\bar{\bf u}}^\prime
e^{-l/2})^2$ $+({\bf u}\!-\!{\bf u}^\prime e^{-l/2})\cdot (\bar{\bf
u}\!-\!{\bar{\bf u}}^\prime e^{-l/2})]/3I^\prime T(1-e^{-l})\}$ and
obtain the three-replica term $S_l({\bf u},\bar{\bf u})$ by simple
integration.  Substituting $S_l$ back into (\ref{frg}) and integrating
over ${\bf u}$ we find that the three-replica terms mutually cancel
and thus do not contribute to the flow of $g_l$.  The amplitude $g_l$
then is determined by the remaining non-linear terms alone,
\begin{equation}
\partial_l g_l=g_l^2/8,
\label{simplified}
\end{equation}
with the solution $g_l=g_0/(1-g_0 l/8)$ as derived from the initial
condition $g_0=(\pi/4) (\epsilon\Delta/T^3)$. At $l_c=8/g_0$ the
amplitude $g_{l_c} \approx 1$ and we find the pinning length
\begin{equation}
L_c(T)\sim\exp[(32/\pi)(T^3/T_{\rm dp}^3)], 
\qquad T_{\rm dp}^3 = \epsilon\Delta,
\label{mainres}
\end{equation} 
the main result of the paper \cite{comment}: the one-loop FRG analysis
allows us to fix the coefficient $C$ in front of the parameter $1/g_0$
in the exponent. Although the criterion $g_{l_c}\simeq 1$ might seem
arbitrary it is sufficient to find the correct asymptotic value of the
RG variable $l_c$ as long as the initial value $g_0$ is small. Indeed,
the solution $g_l$ remains small in most of the interval $[0,l_c]$ and
grows rapidly only very close to $l_c$.

An interesting alternative is offered by mapping the random polymer
problem to the Kardar-Parisi-Zhang equation \cite{KPZ}, where the
dynamic renormalization group (DRG) \cite{FNS} provides excellent
results for the $n=1$ problem \cite{MHKZ}. The weight ${\tilde g}$ in
the (white noise) Ansatz $K_{{\rm wn}} = {\tilde g} \delta^2 ({\bf
u})$ replacing (\ref{simplef}) obeys the equation \cite{Frey}
$\partial_{l^\prime} {\tilde g} = (2-n){\tilde g}+[2(2n-3)/n] {\tilde
g}^2$, with ${\tilde g}(0) =(A_n/(2\pi)^n)\epsilon\Delta/T^3$ (the
prime indicates scaling along the transverse dimension).  For $n=2$
the solution `explodes' at $l_c^\prime = 2\pi(T/T_{\rm dp})^3$,
implying that $L_c(T) \sim \exp[4\pi(T/T_{\rm dp})^3]$ \cite{FH},
different from Eq.~(\ref{mainres}).  Note that, in contrast to the
FRG, the DRG scheme does not renormalize the correlator {\it
function}.  Here, we argue in favor of our result (\ref{mainres}):
While the DRG technique produces accurate results (including a random
manifold fixed point) in dimensions $n<3/2$, the results are less
convincing in larger dimensions $n>3/2$, where the fixed point gives
way to a divergence.  On the other hand, the FRG technique features
both an apparent divergence at $l_c$ and a random manifold fixed point
as $l \rightarrow \infty$ in all dimensions $n \leq 2$, allowing for a
consistent physical interpretation of the results in terms of the
diagram sketched in Fig.\ 1.

Above we have assumed that the integral $\int d^n u \, K_0(u)$
converges, however, flux lines in superconductors exhibit a long range
interaction with the disorder potential spoiling this assumption: for
values $u$ exceeding the coherence length $\xi$, the correlator obeys
the asymptotics $K_0(u)\sim K_0(0)(\xi^2/u^2) \ln (u/\xi)$ and the
integral $\int_{u<u^\ast}d^n u \,K_0(u)$ diverges as $K_0(0)\xi^2
\ln^2({u^\ast}/\xi)$. We estimate $L_c(T)$ for this case: Neglecting
the non-linear term in the first stage of the RG transformation, we
approximate the solution of the Fokker-Planck equation in the region
$u^2 \alt T\xi/\epsilon$ by $[\Delta_l/4\pi T^\prime]
\exp[-u^2/4T^\prime]$ [c.f., Eq.~(\ref{incondnn})], with an
$l$-dependent function $\Delta_l$, a consequence of the divergence of
the integral $\int d^n u\,K_0(u)$. The weight $\Delta_l$ collects
those particles reaching the central region on scale $l$: a
displacement $u^\ast\gg\sqrt{T\xi/\epsilon}$ of the harmonic
oscillator takes a `time' $l^\ast\sim\ln(u^\ast
/\sqrt{T\xi/\epsilon})$ to reach the region $u \alt
\sqrt{T\xi/\epsilon}$ and hence $\Delta_{l^\ast} \sim K_0(0)\xi^2
{l^\ast}^2$. The dimensionless coupling $g_l$ then grows already in
the linear regime due to the `pumping' from large distances. The
pinning length follows from the condition $g_{l_c} =
\epsilon\Delta_{l_c}/T^3 \approx 1$ and we obtain
$L_c(T)\sim\exp[C(T^3/\epsilon K(0)\xi^2)^{1/2}]$ with $C$ of order
unity. The same result follows from a perturbative treatment.

In order to find the prefactor $A(T)$ in the expression (\ref{mainres})
for $L_c(T)$ we have to account for ${\cal O}(\Delta^3)$ corrections, 
resulting in a flow $\partial_l g_l=g_l^2/8 + \eta g_l^3$ with $\eta>0$ 
a constant. This equation explodes at $l_c =8/g_0-64 \eta \ln g_0$,
producing a prefactor of the form $A(T) \simeq \Lambda^{-1}(T_{\rm
dp}/T)^{192\eta}$.  While the two-loop contribution $\eta_{\rm 2-l} =
275/5184$ to $\eta$ can be found from \cite{Bucheli}, the part arising
from the four-replica term is more difficult to obtain --- whether the
four-replica terms mutually cancel (e.g., due to some symmetry), as
was the case in the one-loop analysis above, remains an interesting
and open problem.

We thank L.\ Balents, D.S.\ Fisher, E.\ Frey, V.B.\ Geshkenbein, T.\
Giamarchi, Th.\ Hwa, O.\ Narayan, V.M.\ Vinokur, and J.\ Zinn-Justin
for helpful discussions.


\vspace{-0.5 truecm}
\begin{thebibliography}{99}
\vspace{-1.3 truecm}

\bibitem{Halpin} T.\ Halpin-Healy and Y.-C.\ Zhang, 
Phys.\ Rep.\ {\bf 254}, 215 (1995).

\bibitem{Blatter} G.\ Blatter {\it et al.}, 
Rev.\ Mod.\ Phys.\ {\bf 66}, 1125 (1994).

\bibitem{IV} L.B.\ Ioffe and V.M.\ Vinokur,
J.\ Phys. C {\bf 20}, 6149 (1987).

\bibitem{Gruner} G.\ Gr\"uner,
Rev.\ Mod.\ Phys.\ {\bf 60}, 1129 (1988).

\bibitem{Larkin} A.\ I.\ Larkin,
Zh.\ Eksp.\ Teor.\ Fiz.\ {\bf 58}, 1466 (1970)
[Sov.\ Phys.\ JETP {\bf 31}, 784 (1970)].

\bibitem{HH} D.A.\ Huse and C.L.\ Henley,
Phys.\ Rev.\ Lett.\ {\bf 54}, 2708 (1985).

\bibitem{HHF} D.A.\ Huse {\it et al.}, 
Phys.\ Rev.\ Lett.\ {\bf 55}, 2924 (1985).

\bibitem{Fisher} D.S.\ Fisher, 
Phys.\ Rev.\ Lett.\ {\bf 56}, 1964 (1986).

\bibitem{AV} T.\ Ala-Nissila and O.\ Ven\"al\"ainen, 
J.\ Stat.\ Phys.\ {\bf 76}, 1083 (1994).

\bibitem{FV} M.V.\ Feigel'man and V.M.\ Vinokur, 
Phys.\ Rev.\ B {\bf 41}, 8986 (1990).

\bibitem{Imbrie} J.Z.\ Imbrie and T.\ Spencer, 
J.\ Stat.\ Phys.\ {\bf 52}, 609 (1988).

\bibitem{GB} D.A.\ Gorokhov and G.\ Blatter,
Phys.\ Rev.\ Lett.\ {\bf 82}, 2705 (1999).

\bibitem{Bucheli} H.\ Bucheli {\it et al.}, 
Phys.\ Rev.\ B {\bf 57}, 7642 (1998);
D.A.\ Gorokhov and G.\ Blatter, 
Phys.\ Rev.\ B {\bf 59}, 32 (1999).

\bibitem{Balents} L.\ Balents and D.S.\ Fisher,
Phys.\ Rev.\ B {\bf 48}, 5949 (1993).

\bibitem{GiamarchiLeDoussal} T.\ Giamarchi and P.\ LeDoussal,
Phys.\ Rev.\ B {\bf 52}, 1242 (1995).

\bibitem{comment} Previous results $L_c(T) \propto T^4$ for $n=1$ and
$\ln L_c(T) \propto T^2$ for $n=2$ found by Kim {\it et al.}  [Phys.\
Rev.\ A {\bf 44}, 4782 (1991)] disagree with those from perturbation
theory. Furthermore, the result $L_c(T) \propto \exp[16\pi(T/T_{\rm
dp})^3]$ by Lebedev [Europhys.\ Lett.\ {\bf 16}, 1 (1991)] is based on
the summation of the main terms in the perturbation series, which,
however, have been chosen on the basis of a simplified rule.

\bibitem{KPZ} M.\ Kardar {\it et al.}, 
Phys.\ Rev.\ Lett.\ {\bf 56}, 889 (1986).

\bibitem{FNS} D.\ Forster {\it et al.},
Phys.\ Rev.\ A {\bf 16}, 732 (1977).

\bibitem{MHKZ} E.\ Medina {\it et al.},
Phys.\ Rev.\ A {\bf 39}, 3053 (1989).

\bibitem{Frey} E.\ Frey and U.C.\ Tauber,
Phys.\ Rev.\ E {\bf 50}, 1024 (1994).

\bibitem{FH} D.\ Fisher and D.A.\ Huse,
Phys.\ Rev. B {\bf 43}, 10728 (1991).

\end{thebibliography}
\end{document}